\documentclass[10pt,a4paper,twoside]{article}
\usepackage{epsfig}
\usepackage{baltlat6}
\usepackage{array}
\usepackage{here}
\usepackage{amsmath}

\begin{document}
\ \
\vspace{0.5mm}
\setcounter{page}{53}
\setcounter{footnote}{1}

\titlehead{Baltic Astronomy, vol.\,25, 53--59, 2016}

\titleb{ST\"ACKEL-TYPE DYNAMIC MODEL FOR THE GALAXY BASED ON MASER KINEMATIC DATA}

\begin{authorl}
\authorb{A. O. Gromov}{1},
\authorb{I. I. Nikiforov}{2} and
\authorb{L. P. Ossipkov\footnote{Deceased on August 22nd, 2015.}}{1}
\end{authorl}

\begin{addressl}
\addressb{1}{Department of Space Technologies and Applied Astrodynamics,\\
St.~Petersburg State University, Universitetskij pr.~35,
Staryj Peterhof,\\ St.~Petersburg 198504, Russia; granat08@yandex.ru}
\addressb{2}{Department of Celestial Mechanics, St.~Petersburg State University,\\
Universitetskij pr.~28, Staryj Peterhof, St.~Petersburg 198504, Russia;\\ nii@astro.spbu.ru}
\end{addressl}

\submitb{Received: 2016 January 14; accepted: 2016 March 4}

\begin{summary}
A dynamic model of the Galaxy is constructed based on kinematic data
for masers with trigonometric parallaxes. Maser data is used to
compute the model potential in the Galactic plane. The potential is
then generalized to three dimensions assuming the existence of a
third quadratic integral of motion. The resulting Galactic model
potential is of St\"ackel's type. The corresponding space density
function is determined from Poisson's equation.
\end{summary}


\resthead{Dynamic model of the Galaxy based on maser kinematic data}
{A. O. Gromov, I. I. Nikiforov, L. P. Ossipkov}

\sectionb{1}{INTRODUCTION}

Constructing models for the Galaxy that are based on the data for
masers with trigonometric parallaxes is a popular direction of
research (e.g., Reid et al.\ 2009, 2014; Bajkova \& Bobylev 2015;
Nikiforov \& Veselova 2015). The main advantage of trigonometric
parallaxes is that they determine {\em absolute\/} (geometric)
distances to objects with no assumptions about the distance scale,
luminosity calibration, extinction, metallicity, etc. The
possibilities for {\em accurate\/} VLBI measurements of parallaxes
even for {\em distant\/} masers (see Fig.~5 in Nikiforov \& Veselova
2015) make these objects very important tracers for various
investigations of the Milky Way, and stimulate their intensive
observations (VERA, VLBA, EVN and other projects).

In this paper, we use the data for 103 masers as published by Reid
et al.\ (2014). We convert the maser parallaxes, proper motions, and
radial velocities into Galactocentric distances and rotation
velocities (see Appendix). In Section 2 and 3 we fit the rotation
curves of the one- and two-component model potentials, respectively,
to observational data and estimate the model parameters. To
generalize the potential to three dimensions, we use the theory of
St\"ackel's models (Kuzmin 1952, 1956); we then draw equidensities
for both model potentials using the model of mass distribution
obtained from Poisson's equation (Section~4).

\sectionb{2}{ONE-COMPONENT MODEL}

For the one-component model we use the quasi-isothermal potential
\begin{equation}
\label{Phi}
\Phi(R,0)=\Phi_0^1\ln\left[1+\frac\beta{w(R)}\right],
\end{equation}
where $\beta\in[0,+\infty)$ is a structural parameter of the model,
\begin{equation}
w^2(R)=1+\kappa^2 R^2,
\end{equation}
and $\Phi_0^1$ and $\kappa$ are scale parameters. This potential was
proposed by Kuzmin et al.\ (1986) for spherical systems.

To construct the model for our Galaxy it is necessary to estimate
the parameters $\Phi_0^1$, $\beta$, and $\kappa$. We fit model
circular velocities to the observational data on the Milky Way's
rotation curve. The formula for circular velocity is
\begin{equation}
\label{TC}
\Theta_{\text{c}}^2(R)=-R\frac{\partial\Phi}{\partial R}(R,0).
\end{equation}

We estimate the model parameters by ordinary least-squares fitting.
We minimize the statistic
\begin{equation}
\label{L2}
L^2=\sum_{i=1}^{103}p_i\left[\Theta_{\text{c}}(R_i)-\Theta_i\right]^2,
\end{equation}
where $\Theta_{\text{c}}(R_i)$ is the model circular velocity at
$R_i$ calculated by Eq.~(\ref{TC}); $\Theta_i$ is the ``observed''
rotation velocity calculated from the parallax, proper motion, and
radial velocity of a maser; $p_i=1/{\sigma^{2}_{\Theta_i}}$ is the
weight, and $\sigma^2_{\Theta_i}$ is the measurement error (see
Appendix).

We found that $L^2$ reaches its minimum at $\Phi_0^1=295.4\pm1.4$
km$^2$\,s$^{-2}$, $\kappa=0.4346\pm0.0057$~kpc$^{-1}$, and
$\displaystyle q=0.9002\pm0.0014$. The quasi-isothermal model with
these parameter values provides the best approximation to
observational data.

The left panel in Fig.~1 compares the model rotation curve with
observational data. Here the solid curve, dots, and vertical bars
show the model velocity curve $\Theta_\text{c}(R)$,  maser data, and
the $\Theta_i$\, measurement errors, respectively. The mean error of
unit weight for this solution is $\sigma\equiv
L/\sqrt{N_{\text{free}}}=3.2$.  Large
$\sigma=\sqrt{\chi^2/\text{DOF}}\gg 1$ means that residuals can not
be explained by measurement errors.

Earlier we constructed a similar model by fitting the same model
rotation curve to six independent H\,I data sets. From these data we
found $\Phi_0^1=258.1\pm1.5$ km$^2$\,s$^{-2}$,
$\kappa=0.3202\pm0.0052$~kpc$^{-1}$, and $\displaystyle
q=1^{+0}_{-0.008}$ (Gromov et al.\ 2015). It is the limiting case of
the quasi-isothermal model, i.e., the so-called Jaffe model. The
mean error of unit weight in this case is $\sigma=2.98$
km\,s$^{-1}$. Note that we set the weights for H\,I data points
proportionally to the length of interval of Galactocentric distances
$[x_{\text{min}}, x_{\text{max}}]$ covered by the respective data
set (see Gromov et al.\ 2015 and reference therein). Here, $x=R/R_0$
and $R_0$ is the solar Galactocentric  distance. Thus $L^2$ and
$\sigma$ for H\,I data are dimensional statistics, whereas the
corresponding functions for masers are dimensionless. Fig.~1
compares the rotation curves constructed for maser (the left panel)
and H\,I (the right panel) data.

\sectionb{3}{TWO-COMPONENT MODEL}

Multi-component models usually agree better with observational data.
Furthermore, the Galaxy has a multi-component structure and
therefore each component should be described by its own model
potential. We consider a two-component model with the potential

\begin{equation}
\Phi=\Phi_1+\Phi_2\,,
\end{equation}
where $\Phi_1$ is quasi-isothermal potential (\ref{Phi}) and
$\Phi_2$ is the generalized-isochrone potential
\begin{equation}
\Phi_2=\Phi_0^2\frac{\alpha}{\left(\alpha-1\right)+\sqrt{1+\kappa_1^2 R^2}}\,.
\end{equation}

\begin{figure}[!tH]
\vbox{ \psfig{figure=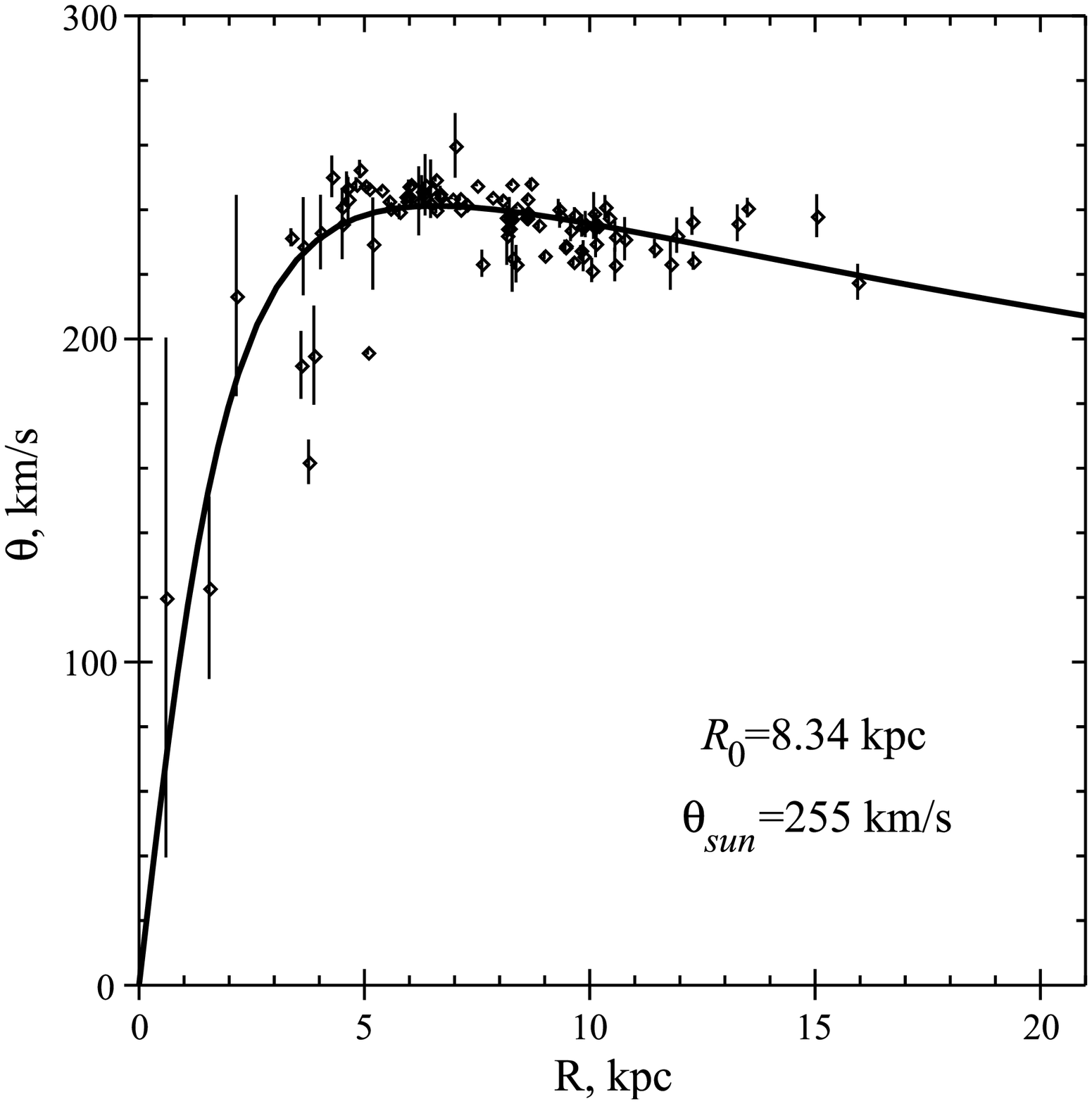,width=60mm,angle=0,clip=} \hspace{2mm}
\psfig{figure=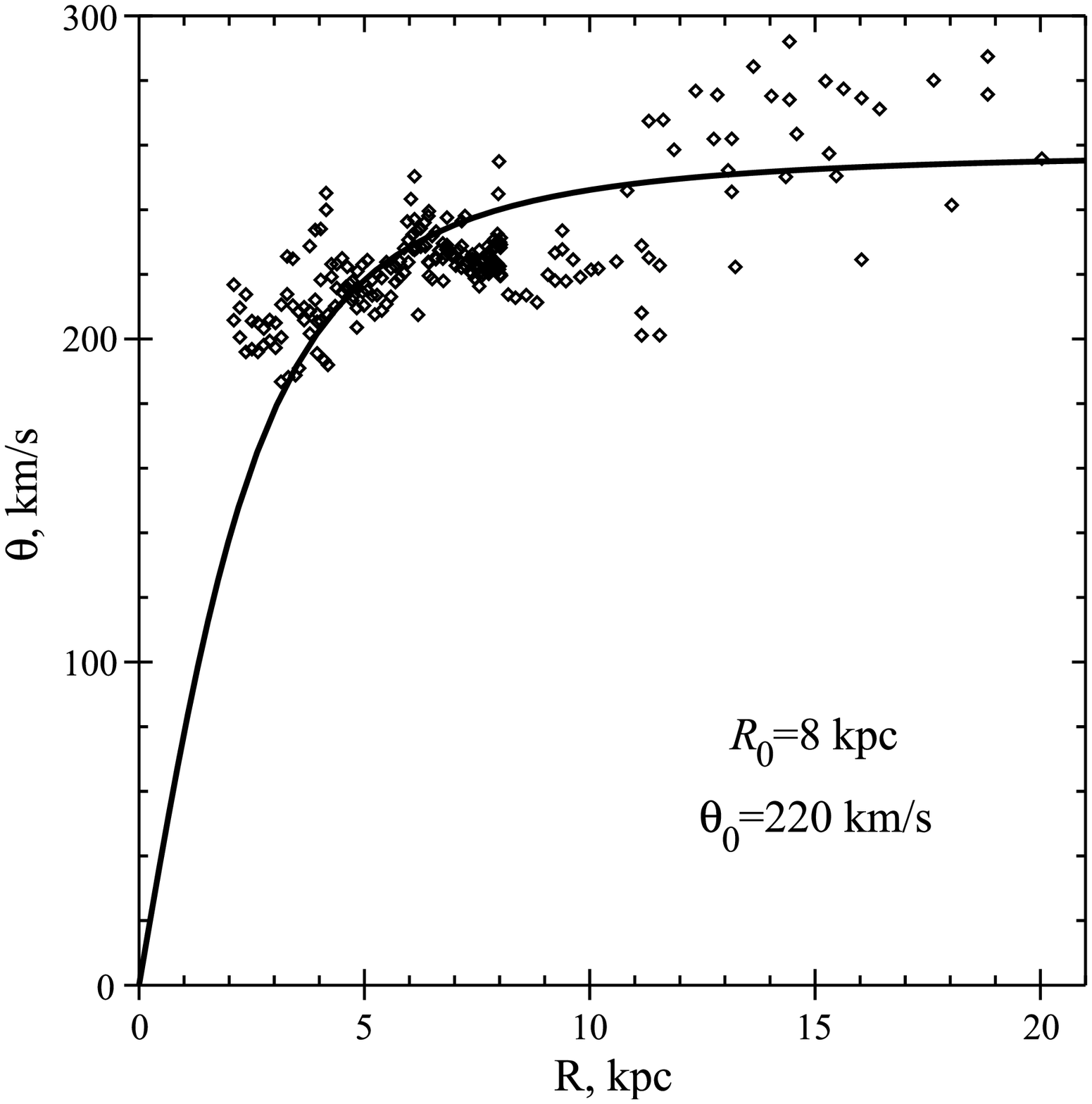,width=60mm,angle=0,clip=} \vspace{1mm}
\captionb{1} {Comparison of rotation curves for the one-component
model constructed for maser (left) and H\,I (right) data. } }
\end{figure}

Minimizing the function $L^2$ computed for the two-component model
applied to maser data yields $\kappa~=~0.701\pm0.047$~kpc$^{-1}$,
$q=0.99233\pm0.00084$, $\Phi_0^1=228.0\pm1.3$ km$^2$\,s$^{-2}$,
$\alpha=1.41\pm0.12$, $\kappa_1=0.1467\pm0.0055$~kpc$^{-1}$, and
$\Phi_0^2=178.4\pm4.5$~km$^2$\,s$^{-2}$. We compare the
corresponding rotation curve with observational data in Fig.~2 (the
left panel). The mean error of unit weight is $\sigma=2.8$, i.e.,
smaller than in the case of the one-component model, and hence the
data are better described by the two-component model. However,
$\sigma$ is still much greater than unity.

Our analysis of H\,I data yields
$\kappa=0.07379\pm0.00051$~kpc$^{-1}$, $q=0.9427\pm0.0078$,
$\Phi_0^1=336.3\pm5.9$ km$^2$\,s$^{-2}$, $\alpha=0.403\pm0.023$,
$\kappa_1=0.0574\pm0.0037$~kpc$^{-1}$, and
$\Phi_0^2=288.2\pm5.4$~km$^2$\,s$^{-2}$ with a mean unit weight
error of $\sigma=2.44$ km\,s$^{-1}$ (Gromov \& Nikiforov 2015). We
compare the corresponding rotation curve  with H\,I data in Fig.~2
(the right panel).

\begin{figure}[!tH]
\vbox{ \psfig{figure=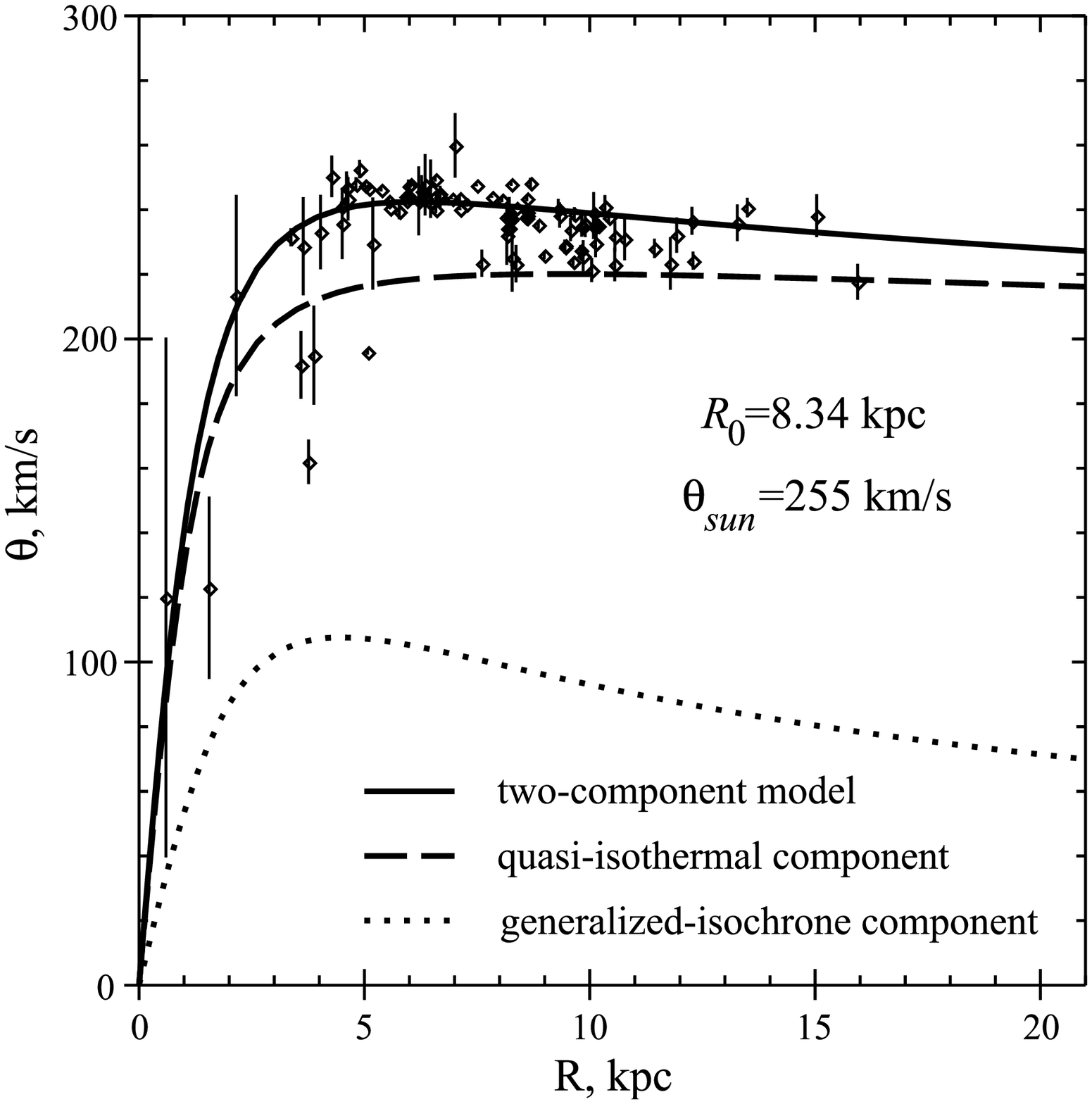,width=60mm,angle=0,clip=} \hspace{2mm}
\psfig{figure=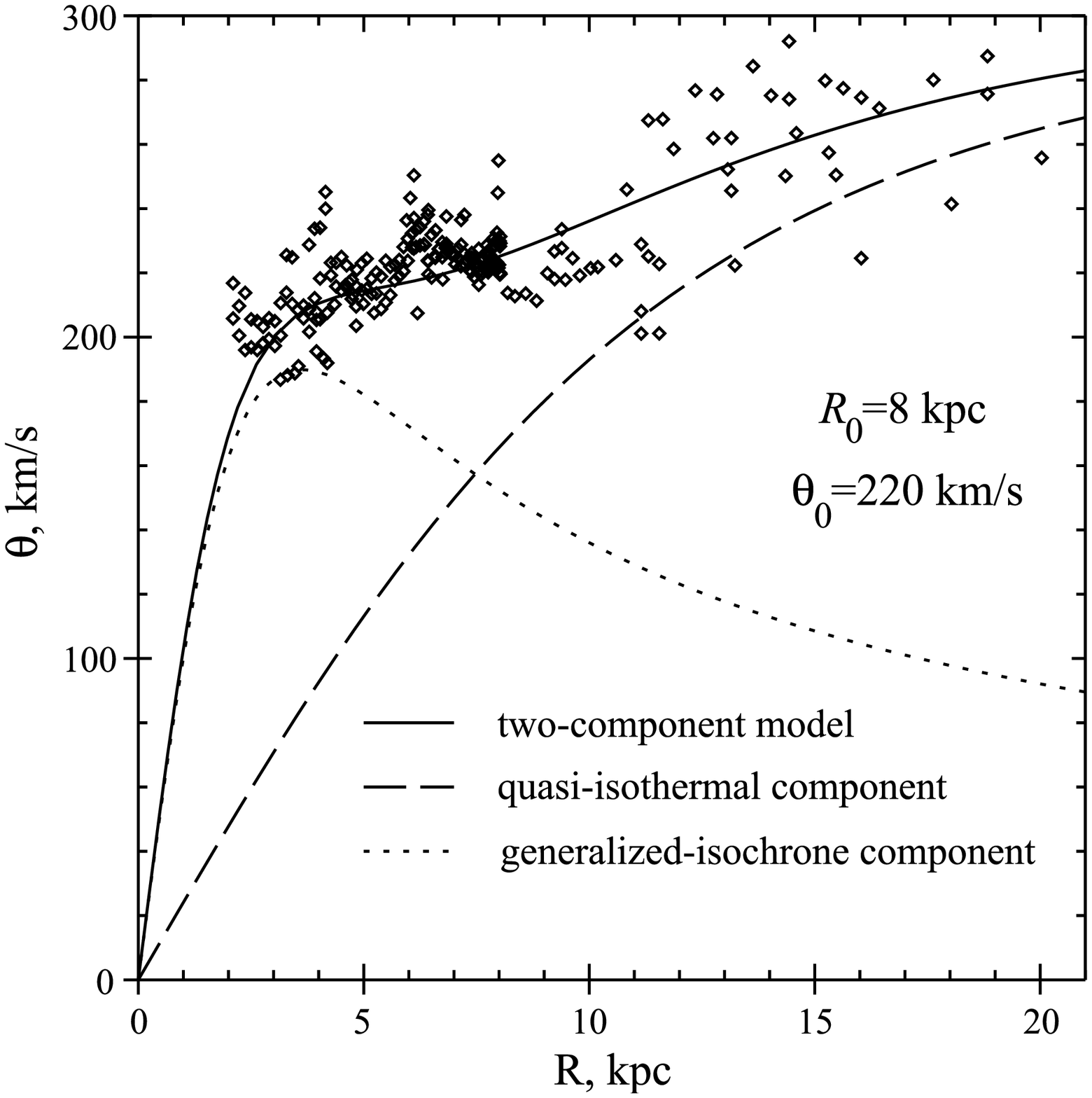,width=60mm,angle=0,clip=} \vspace{1mm}
\captionb{2} {Comparison of rotation curves for the two-component
model constructed from maser (left) and H\,I (right) data. } }
\end{figure}

\sectionb{4}{GENERALIZATION OF POTENTIAL TO THREE DIMENSIONS}


We use the theory of St\"ackel's potentials (Kuzmin 1952, 1956) to
generalize the derived potential to three dimensions. We assume that
a third integral of motion exists that depends quadratically on
velocities:
\begin{equation}
\label{I3}
I_3=(R\,v_z-z\,v_R)^2+z^2v_{\theta}^2+z_0^2(v_z^2-2\Phi^*)\,,
\end{equation}
where $z_0$ is a scale parameter of dimension of
length, and function $\Phi^*(R,z)$ must satisfy the equations
\begin{equation}
\displaystyle
\displaystyle z_0^2\frac{\partial\Phi^*}{\partial R}=z^2\frac{\partial\Phi}{\partial R}-Rz\frac{\partial\Phi}{\partial z}\,,\qquad
\linebreak
\displaystyle z_0^2\frac{\partial\Phi^*}{\partial z}=(R^2+z_0^2)\frac{\partial\Phi}{\partial z}-Rz\frac{\partial\Phi}{\partial R}\,.
\end{equation}

In the elliptic coordinates
$\xi_1\in[1;\infty)$, $\xi_2\in[-1;1]$,
\begin{equation}
R=z_0\sqrt{\left(\xi^2_1-1\right)\left(1-\xi^2_2\right)}\,, \qquad
z=z_0\,\xi_1\,\xi_2\,,
\end{equation}
St\"ackel's potentials have the following form:
\begin{equation}
\Phi=\frac{\varphi(\xi_1)-\varphi(\xi_2)}{\xi^2_1-\xi^2_2}\,,
\end{equation}
where $\varphi(\xi)$ is an arbitrary function. Determining function
$\varphi(\xi)$ for some potential means generalizing this potential
to 3D space. To find $\varphi(\xi)$, we use formulas derived by
Rodionov (1974).

For our one-component model,
\begin{equation}
\varphi(\xi)=\xi^2\Phi_0^1\ln\left(1+\frac\beta
{\sqrt{1+\kappa^2z_0^2(\xi^2-1)}}\right)\,
\end{equation}
(Gromov 2013, 2014a), and for the two-component model,
$$
\hspace{-80pt}\varphi(\xi)=\xi^2\Phi_0^1\ln\left(1+\frac\beta
{\sqrt{1+\kappa^2z_0^2(\xi^2-1)}}\right)+{}
$$
\begin{equation}
\hspace{80pt}{}+\xi^2\Phi_0^2\frac{\alpha}{\left(\alpha-1\right)\sqrt{1+\kappa_1^2z_0^2(\xi^2-1) }}\,
\end{equation}
(Gromov 2014b).

We use the spatial density derived  from Poisson's equation (Gromov
2013, 2014a,b) to draw the density contours for both models (Figs.~3
and~4). Here we adopt the parameter values inferred from maser data
(see Sections~2 and~3). The parameter $z_0$, which appears in the
formula for density, is determined from the following equation:
\begin{equation}
 \label{z0}
\displaystyle
z_0^2(R)=\left.\displaystyle\left[\frac{3\displaystyle\frac{\partial \Phi(R,z)}{\partial
R}+R\left(\displaystyle\frac{\partial^2 \Phi(R,z)}{\partial R^2}-4\frac{\partial^2
\Phi(R,z)}{\partial z^2}\right)}{\displaystyle\frac{\partial^3 \Phi(R,z)}{\partial  z ^2
\partial R}}\right]\right|_{z=0}-R^2
\end{equation}
(Ossipkov 1975). Equation~(\ref{z0}) is the constraint that the
third integral of motion imposes on the potential. We assume that in
the solar neighborhood the potential is close to that proposed by
Gardner et al.\ (2011), and substitute the latter into
Equation~(\ref{z0}). Note that the above authors constructed their
potential based on the data on the vertical component of the
Galactic tidal field, and we therefore assume that it should
describe the vertical structure of our Galaxy quite well. For
Gardner et al.'s potential, $z_0$~=~$5.3$~kpc in the solar
neighborhood ($R=8$~kpc).

\begin{figure}[!tH]
\vspace{-3pt} \vbox{
\centerline{\psfig{figure=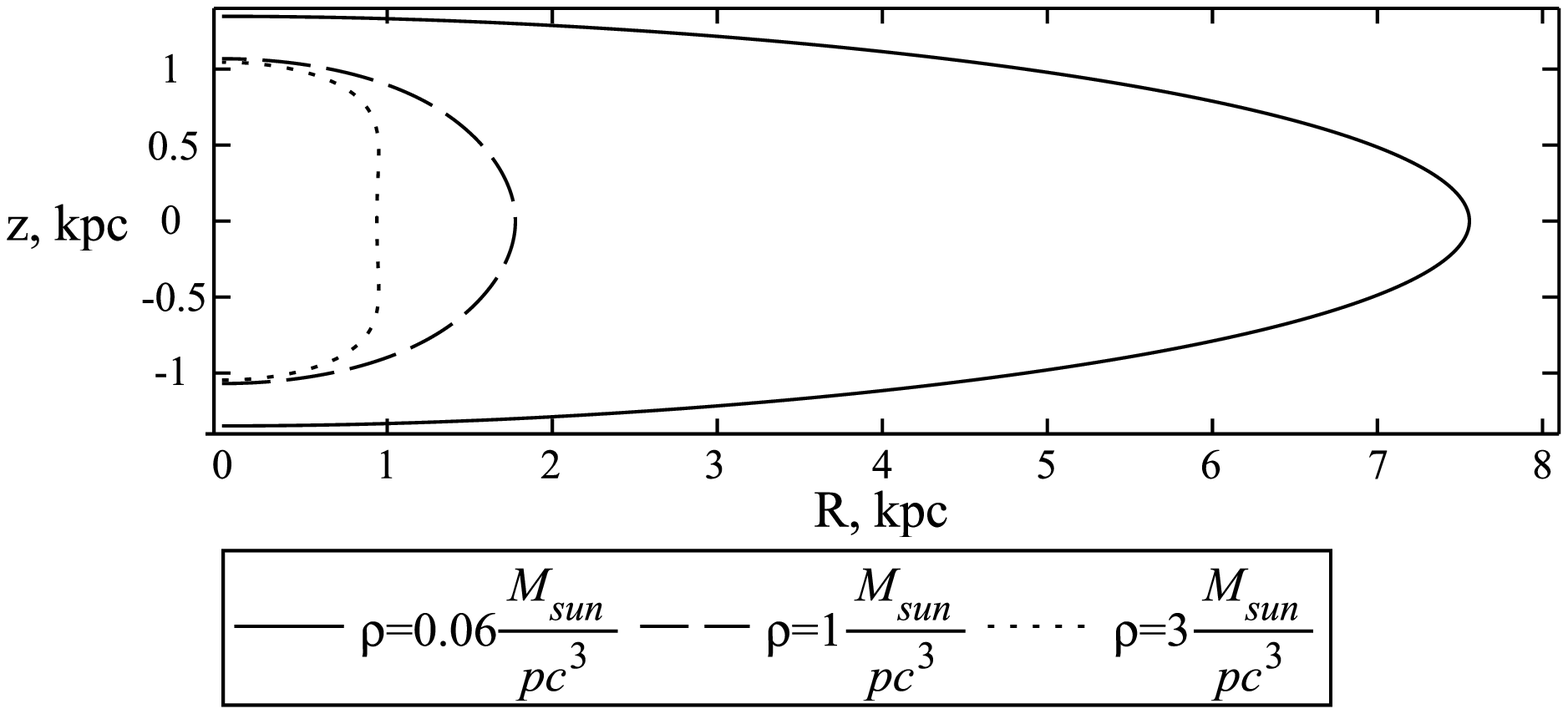,width=90mm,angle=0,clip=}}
\vspace{-8pt} \vspace{1mm} \captionb{3} {Density contours for the
one-component model based on maser data.}\hspace{5mm}}
\end{figure}

\begin{figure}[!tH]
\vbox{
\psfig{figure=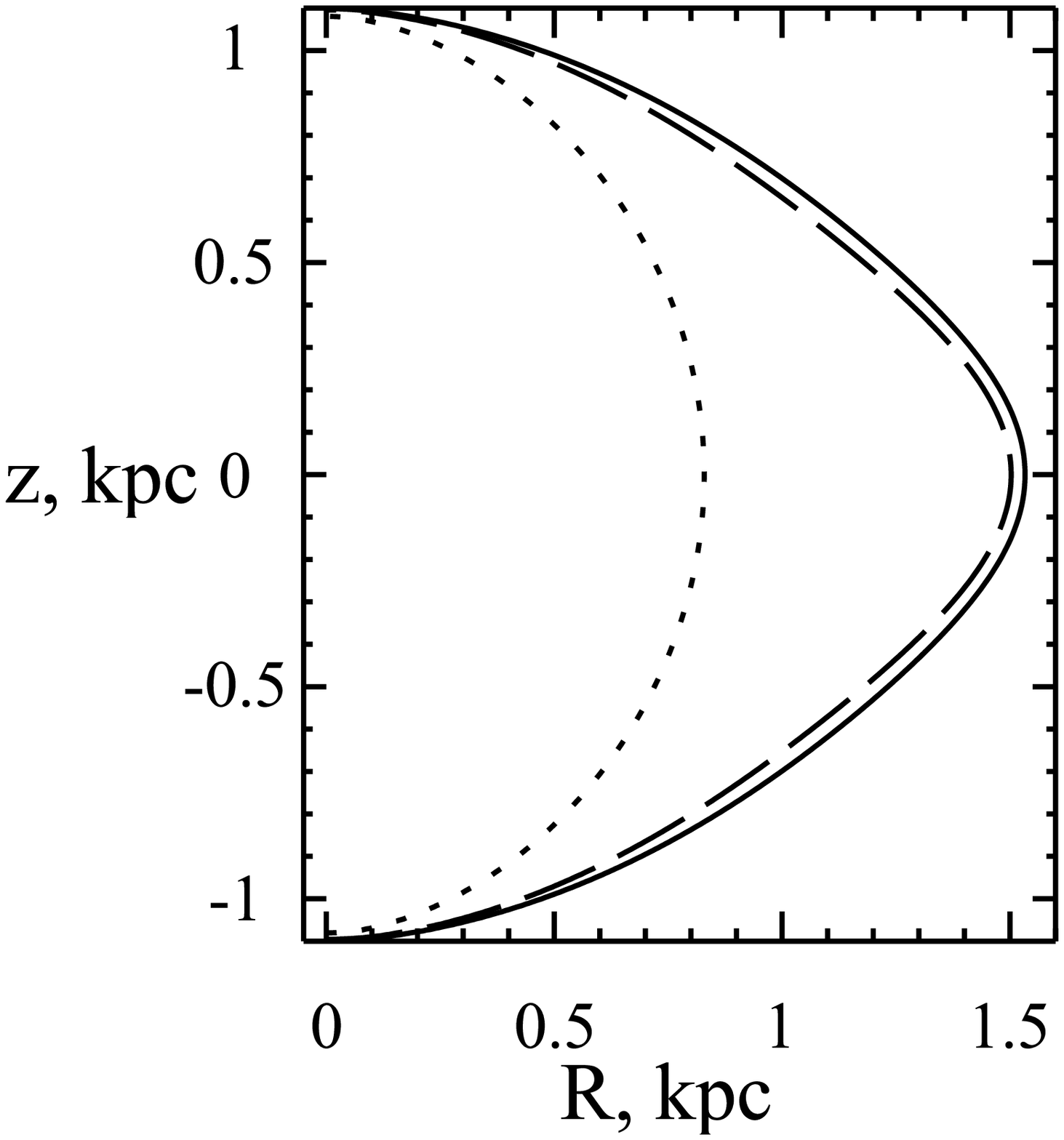,width=45mm,angle=0,clip=}
\raisebox{5pt}{\psfig{figure=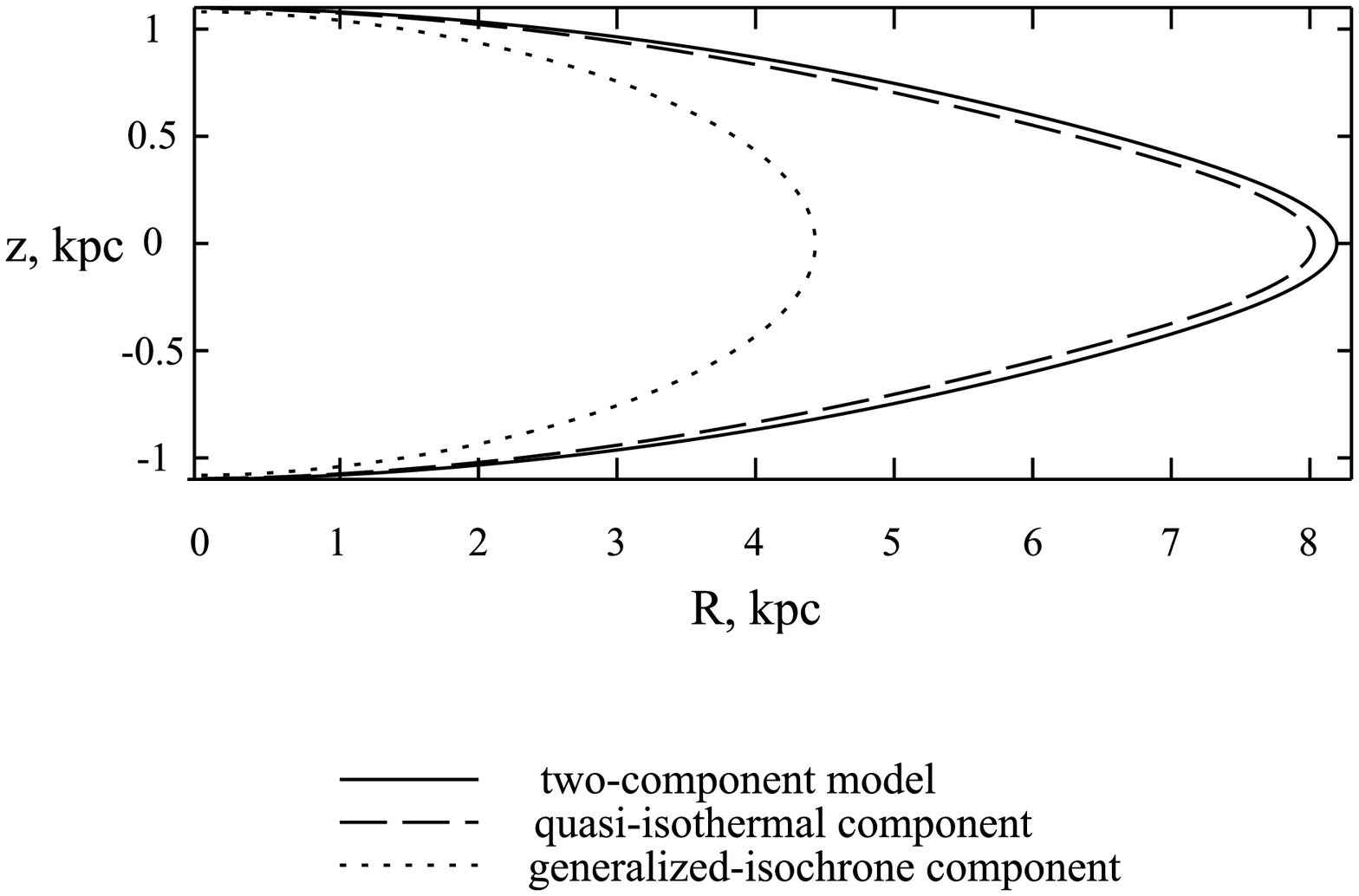,width=77mm,angle=0,clip=}}
\vspace{-1mm} \captionb{4} {Density contours for the two-component
model based on maser data: $ \rho=1~M_{\odot}$\,pc$^{-3}$ (left), $
\rho=0.08~M_{\odot}$\,pc$^{-3}$ (right).} }
\end{figure}

\sectionb{5}{DISCUSSION AND CONCLUSIONS}

We used the observational data for masers to consider the
possibility of applying the quasi-isothermal and two-component
models (with the quasi-isothermal and ge\-ne\-ra\-lized-isochrone
potentials)  to our Galaxy. The models constructed fit the data
well. The corresponding unit-weight errors, $\sigma \approx 3\pm
0.2$, show that a more correct system of weights is needed to
eliminate eventual systematic biases. We plan to introduce such a
system of weights, although it will complicate the procedure of
constructing the model. We also plan to pay attention to the
treatment of outlying data.

It follows from a comparison of the results obtained using maser
data with those based on H\,I observations that the parameter $q$ is
close to unity. Hence the models are similar to the limiting case,
i.e., to the Jaffe model. However, the parameter $q$ for the
two-component models does not reach unity, and hence the
two-component models are more physical. Such  $q$'s result in the
elliptical shape of density contours.

We constructed the model of mass distribution by generalizing the
potential to 3D space using the theory of St\"ackel's potentials.
Note that the model density values in the solar neighborhood,
$\rho=0.06~M_{\odot}$\,pc$^{-3}$, and $0.08~M_{\odot}$\,pc$^{-3}$
for the one- and  two-component models, respectively, are close to
the observed density, $\rho=0.08-0.11~M_{\odot}$\,pc$^{-3}$ (e.g.,
Loktin \& Marsakov 2010).

Physically, it would be natural to construct a three-component model
of the Galaxy representing the halo, disk, and bulge. In our
two-component model the quasi-isothermal component represents the
disk and halo, and the ge\-ne\-ra\-lized-isochrone component, the
bulge.

\thanks{One of us (IIN) acknowledges the support from Saint Petersburg State University
(grant No.~6.37.341.2015).}

\References

\refb Bajkova A. T., Bobylev V. V. 2015, Baltic Astronomy, 24, 43

\refb Gardner E., Nurmi P., Flynn C., Mikkola S. 2011, MNRAS, 411, 947

\refb Gromov A. O. 2013, Izv.\ Glavn.\ Astron.\ Obs. (Pulkovo), 221, 129

\refb Gromov A. O. 2014a, Vest.\ Saint Petersburg Univ., ser. 1, 2, 322

\refb Gromov A. O. 2014b,  Astron.\ and Astrophys.\ Trans., 28, 331

\refb Gromov A. O., Nikiforov I. I. 2015, Izv.\ Glavn.\ Astron.\
Obs. (Pulkovo), 222, 31

\refb Gromov A. O., Nikiforov I. I., Ossipkov L. P. 2015, Baltic Astronomy, 24, 150

\refb Kuzmin G. G. 1952, Publ.\ Tartu Obs., 32, 332

\refb Kuzmin G. G. 1956, AZh, 33, 27

\refb Kuzmin G. G., Veltmann \"U.-I. K.,  Tenjes P. L. 1986, Publ.\ Tartu Obs., 51, 232

\refb Loktin A. V., Marsakov V. A. 2010, Lectures on Stellar Astronomy, Rostov-na-Donu,
pp. 282 (in Russian)

\refb Nikiforov I. I., Veselova A. V. 2015, Baltic Astronomy, 24,
387

\refb Ossipkov L. P. 1975, Vest. Leningrad Univ., 7, 151

\refb Rodionov V. I. 1974, Vest. Leningrad Univ., 13, 142

\refb Reid M. J., Menten K. M., Zheng X. W. et al. 2009, ApJ, 700,
137

\refb Reid M. J., Menten K. M., Brunthaler A. et al. 2014, ApJ, 793,
72

\vspace{5mm} APPENDIX\nopagebreak
\vspace{1mm} \noindent\baselineskip=11pt

We basically follow the procedure outlined by Reid et al.\ (2009).
\begin{enumerate}
\item Conversion of the velocity relative the LSR, $V_{\text{lsr}}$, into heliocentric
velocity~$V_r$:
$$ V_r=V_{\text{lsr}}-U_{\odot}\cos l \cos b-V_{\odot} \sin l \cos b-W_{\odot}
\sin b\,, $$ where $U_{\odot}=10.3$ km\,s$^{-1}$, $V_{\odot}=15.3$ km\,s$^{-1}$,
$W_{\odot}=7.7$ km\,s$^{-1}$, according to a value of 20~km\,s$^{-1}$ toward
$\alpha(1900)=18^{\text{h}}$, $\delta(1900)=+30^\circ$.

\item Conversion of equatorial coordinates $(\alpha,\delta)$ into the galactic
coordinates~$(l,b)$:
$$
\sin b= \sin\delta\cos(90^{\circ}-\delta_{\text{p}})-\cos\delta\sin(\alpha-\alpha_{\text{p}}-6^{\text{h}})\sin(90^{\circ}-\delta_{\text{p}})\,,
$$
$$
\hspace{-10pt}\sin\varphi=\left[\cos\delta\sin(\alpha-\alpha_{\text{p}}-6^{\text{h}})\cos(90^{\circ}-\delta_{\text{p}})+\sin\delta\sin(90^{\circ}-\delta_{\text{p}})\right]\slash\cos b\,,
$$
$$
\cos\varphi=\cos\delta\cos(\alpha-\alpha_{\text{p}}-6^{\text{h}})\slash\cos b\,, \qquad l=\varphi+(\theta-90^{\circ})\,,
$$ where $\alpha_{\text{p}}=12^{\text{h}}51^{\text{m}}26\fs 2817$,
$\delta_{\text{p}}=27^{\circ}07'42\farcs 013$, $\theta=122\fdg 932$ 

\item Conversion of the proper-motion components in equatorial coordinates ($\mu_{\alpha},
\mu_{\delta}$) into the motion components in Galactic coordinates
($\mu_l, \mu_b$): $$
\mu_l=l(\alpha+\mu_{\alpha},\delta+\mu_{\delta})-l(\alpha,\delta)\,,
\qquad
\mu_b=b(\alpha+\mu_{\alpha},\delta+\mu_{\delta})-b(\alpha,\delta)\,.
$$

\item The calculation of the velocity components:
$$
 V_l=k\,r\mu_l\cos b\,, \qquad V_b=k\,r\mu_b\,,
$$

where $k=4.7406$.

\item Conversion to the Cartesian heliocentric coordinate system: $$ U=(V_r\cos b-V_b\sin b)\cos
l-V_l
\sin l, $$ $$ V=(V_r\cos b-V_b\sin b)\sin l+V_l\cos l, $$ $$ W=V_b\cos b+V_r\sin b.
$$

\item Conversion to the Galactocentric coordinate system associated with the Sun: $$
U_{\text{g}}=U+U_{\odot}\,, \qquad V_{\text{g}}=V+\theta_{\odot}\,, \qquad
W_{\text{g}}=W+W_{\odot}\,, $$ where $U_{\odot}=11$ km\,s$^{-1}$, $\theta_{\odot}=255$
km\,s$^{-1}$, $W_{\odot}=9$ km\,s$^{-1}$ (Reid et al.\ 2014).

\item Conversion in the Galactocentric coordinate system associated with the object: $$
R^2=R_0^2+r^2\cos^2 b-2R_0r\cos l \cos b, \qquad
\Theta=V_{\text{g}}\cos\beta+U_{\text{g}}\sin\beta, $$
$$\sin\beta=\frac{r\cos\beta}{R}\sin l,\qquad
\cos\beta=\frac{R_0-r\cos b\cos l}{R}\,,$$
where $R_0=8.34$ kpc  (Reid et
al.\ 2014).

\item The measurement error in $\Theta$  is
$$
\sigma^2_{\Theta}=\left(\frac{\partial \Theta}{\partial \pi}\right)^2\sigma^2_{\pi}+\left(\frac{\partial \Theta}{\partial \mu_{\alpha}}\right)^2\sigma^2_{\mu_{\alpha}}+\left(\frac{\partial \Theta}{\partial \mu_{\delta}}\right)^2\sigma^2_{\mu_{\delta}}+\left(\frac{\partial \Theta}{\partial V_r}\right)^2\sigma^2_{{\text{V}}_r}\,,
$$ where the measurement errors $\sigma_{\pi}$, $\sigma_{\mu_{\alpha}}$,
$\sigma_{\mu_{\delta}}$, $\sigma_{{\text{V}}_r}$ are adopted from Reid et al.\ (2014).
\end{enumerate}

\end{document}